# Wavelength-sized GaAs optomechanical resonators with GHz frequency

L. Ding[1a)], C. Baker[1], P. Senellart[2], A. Lemaitre[2], S. Ducci[1], G. Leo[1], I. Favero[1b)]

[1]*Laboratoire Matériaux et Phénomènes Quantiques, Université Paris Diderot, CNRS, UMR 7162, 10 rue Alice Domon et Léonie Duquet, 75013 Paris, France*

[2]*Laboratoire de Photonique et Nanostructures, CNRS, Route de Nozay, 91460 Marcoussis, France*

We report on wavelength-sized GaAs optomechanical disk resonators showing ultra-strong optomechanical interaction. We observe optical transduction of a disk mechanical breathing mode with 1.4 GHz frequency and effective mass of ~ 2 pg. The measured vacuum optomechanical coupling rate reaches $g_0$ = 0.8 MHz, with a related differential optomechanical coupling factor $g_{om}$ = 485 GHz/nm. The disk Brownian motion is optically resolved with a sensitivity of $10^{-17}$ m/√Hz at room temperature and pressure.

Optomechanical systems[1-3] combining a mechanical oscillator and an optical cavity find now applications in very different fields of physics, from mesoscopic quantum physics to cold atoms[4-5] and mechanical sensing[6]. GHz mechanical oscillators can help accessing the quantum regime of optomechanics, can allow developing ultra-fast sensing systems, or can match hyperfine transitions of (artificial) atoms interfaced with the mechanical system. The difficulty lies generally in coupling such GHz oscillators to photons efficiently, in order to offer optical control over the oscillator motion, together with fine optical read-out sensitivity.[7] A useful way

---

[a)]Electronic mail: lu.ding@univ-paris-diderot.fr
[b)]Electronic mail: ivan.favero@univ-paris-diderot.fr

to quantify the optomechanical coupling is to use the differential "frequency pull" parameter $g_{om}$ defined as $g_{om} = d\omega_o/d\alpha$ expressing how the deformation of the optical resonator $\alpha$ modifies its optical resonance eigenfrequency $\omega_o$. In a quantum description of the optomechanical interaction at the single photon level, another relevant parameter is the vacuum optomechanical coupling $g_0$ given by $g_{om}x_{ZPF}$, where $x_{ZPF} = \sqrt{\hbar/2m_{eff}\omega_m}$ is the zero-point motion of the mechanical oscillator, $m_{eff}$ its effective motional mass and $\omega_m$ its resonance angular frequency. Large optomechanical coupling, approaching $g_{om} = \omega_o/\lambda$ with $\lambda$ the wavelength of photons, was recently obtained in photonic-crystal based structures.[7-9] In this work, we scale the size of GaAs optomechanical disk resonators[10] down to $\lambda$ and show that we can this way naturally couple an (over) GHz mechanical oscillator to an optical cavity of high quality factor Q and sub-wavelength mode volume. Thanks to the strong confinement of the optical and mechanical modes in a nano-scale disk, we approach $g_{om}^e = \omega_o/\lambda$ and obtain a vacuum coupling rate $g_0^e = 0.8$ MHz.

Our wavelength-sized GaAs disks stand on christmas-tree-like $Al_{0.8}Ga_{0.2}As$ pedestals[11] and are fabricated from an epitaxial wafer using e-beam lithography and two-steps wet etching[12]. The typical disk has a radius R = 1 µm and a thickness of 200 nm, as seen in Fig. 1(a). For such a disk, 2D axis-symmetry finite-element-method (FEM) simulations[13] predict in the 1500-1600 nm range one unique optical whispering gallery mode (WGM) resonance at a wavelength of $\lambda_0 \approx 1550$ nm. The corresponding WGM is a TE (p = 1, m = 7) mode, where p and m are radial and azimuthal numbers respectively.[14] The electric field intensity is localized at the periphery of the disk [Fig 1(b)]. The free spectral range between two consecutive m numbers of this WGM is $\lambda_0^2/(2\pi n_{eff}R) \approx 140$ nm, taking R = 1 µm and $n_{eff}$ = 2.6 for the effective index of the 200 nm GaAs slab TE mode. The mechanical mode of the disk is simulated by 3D FEM simulation. Based on our previous work[10], we know that axis-symmetry breathing modes with azimuthal number M = 0 couple maximally to WGMs photons. For the present disk, a breathing mode with M = 0 is expected at a frequency around 1.5 GHz. Its 3D deformation profile is shown in Fig. 1(c).

Figure 1(d) is a schematics of near-field optomechanical experiments on a single disk[15], where light from a mode-hope free tunable external-cavity diode laser ($\lambda$ = 1500 – 1600 nm, linewidth ~ 150 KHz) is evanescently coupled into the disk using a looped fiber-taper method.[15-16] All experiments and simulations are carried out at a room temperature and pressure.

The optical spectra of three GaAs disks are shown in Figs. 2(a)-(c), at low and high optical power. They are recorded from a weak DC component of the fiber output. At low power, the symmetric optical resonance indicates linear behavior of the disk optical mode. The intrinsic optical Q reaches $1 \times 10^4$, mainly limited at this wavelength by radiation losses. In our experiments, we measure the optical resonance wavelength with an error inferior to 50 pm. Figure 2(d) shows the WGM measured wavelength $\lambda_o$ as a function of the disk radius, the latter being measured in a Scanning Electron Microscope (SEM) with an accuracy of ± 50 nm. Relying on an epitaxially controlled thickness of 200 nm, the measured wavelength relates through numerical simulations to a given disk radius (solid squares), which agrees well with the measured radius (circles with error bar). In our study, the WGM has little dependence on the pedestal geometry.

At high optical power, we observe an important thermo-optic distortion of the optical resonance.[15,17-18] The strong confinement of optical energy in the disk combines with residual absorption and a poor thermal anchoring to produce significant heating. Figure 2(e) shows the relative optical resonance shift $\Delta\lambda$ as a function of the power transmitted at the fiber output. The intracavity circulating power is proportional to this optical power transmitted by the fiber. Red, green, and blue circles represent disk 1, 2, and 3, respectively. Considering a linear thermal-optic effect $n(T) = n(300K) + (dn/dT) \times \Delta T$ and the approximate relation $2\pi R n_{eff} = m\lambda_0$, the temperature increase corresponding to a shift of the resonance wavelength $\Delta\lambda$ [Fig. 2(f)] is $\Delta T = n_{eff}/(dn/dT) \times (\Delta\lambda/\lambda_0)$, where $dn/dT = 2.35 \times 10^{-4}$ K$^{-1}$ in GaAs at room temperature[19]. In extreme cases, the resonance shift reaches 100 nm corresponding to a temperature increase $\Delta T > 700$ K at which the disk is irreversibly damaged after desorption of As atoms from the GaAs surface[20]. Note that heating can also induce optical resonance shift by means of direct thermal expansion of the disk gallery, in proportion of $\alpha_L = 6.05 \times 10^{-6}$ K$^{-1}$, the room temperature thermal expansion coefficient[21]. The associated shift, $\Delta\lambda = \lambda_0 \alpha_L \Delta T$, is however two orders of magnitude smaller than the thermo-optical shift. This contribution is hence negligible and the optical resonance shift observed in our experiments is safely attributed to thermo-optical effects. In Fig. 2(e), the observed approximate linear scaling of the temperature increase with optical power is consistent with the fact that the three disks have similar geometry. However the exact optical absorption and thermal dissipation mechanisms remain to be investigated and possibly optimized for future experiments.

When the laser wavelength is tuned to a flank of the WGM resonance, the disk mechanical motion modulates the transmitted optical power, whose RF noise spectrum becomes a vibrational spectrum of the disk. Because the disk mechanical mode of interest resonates over 1

GHz, we use a high-speed PIN photodetector with 2 GHz bandwidth. We replace electronic amplification of the detector, which would limit its bandwidth, by optical amplification of the signal at the fiber output, using an erbium-doped fiber amplifier [see Fig 1(d)]. The output of the photodetector is then analysed by a spectrum analyzer. Figures 3(a)-(c) show the obtained RF-spectrum of each disk. The mechanical frequency of the breathing mode is found at 1.4 GHz. Using the disk radius inferred from the optical resonance and pedestal geometry observed in the SEM, we simulate numerically the mechanical mode for each disk and obtain a correct agreement with experimental frequencies. We estimate that the irregular shape of the pedestal leads to an error of ± 50 MHz in the simulated value. At high optical power, the optically induced heating barely modifies the material's elasticity so that the associated mechanical frequency shift, about 1%, is negligible.

The optomechanical coupling between the identified optical and mechanical modes is computed using a perturbative treatment of Maxwell's equations[10,22], yielding a simulated optomechanical coupling of $g_{om}^s$ = 690 GHz/nm for disk 1, with a related vacuum coupling of $g_0^s$ = 1.1 MHz (see Table 1 for values on the three disks). The optomechanical coupling can also be measured independently by calibrating the optical measurement knowing that the Brownian motion of the disk is transduced optically in proportion of $g_{om}$.[7,10] For disk 1, this second method yields an experimental optomechanical coupling of $g_0^e$ = 0.8 MHz ($g_{om}^e$ = 485 GHz/nm), a factor 1.4 times smaller than the FEM-estimated value. Similar discrepancies were observed in photonic crystals strucutres[7-9] and probably stem from an imperfect control of the geometry obtained by nanofabrication. Table 1 lists mechanical and optomechanical parameters of the three studied disks. The calibrated Brownian motion of disk 1 is plotted in Fig.

3(d), showing a sensitivity of a few $10^{-17}$ m/√Hz, which is only a factor 100 above the Standard Quantum Limit imprecision[23]. The second disk in the table possesses a larger pedestal than the disk seen in Fig. 1(a). This results in a less confined optical WGM and a reduced effective motional mass. The optomechanical coupling is consequently reduced with respect to the two other disks.

In summary, wavelength-sized GaAs optomechanical disk resonators with mechanical frequency over the GHz are observed. The vacuum optomechanical coupling $g_o$ is measured to rise up to the MHz with associated $g_{om}$ of 500 GHz/nm. This strong optomechanical interaction makes GaAs disks an interesting platform to study quantum-optomechanics at the single photon level[24], especially since solid state single photons emitters in form of InAs quantum dots[25] are already being inserted in such disks in quantum electrodynamics experiments[26,27].

This work was supported by C-Nano Ile de France.

Table 1 Experimental (superscript e) and simulated (superscript s) mechanical and optomechanical parameters of three miniature GaAs disks.

| $R_{disk}$ (μm) | $f_M^e$ (GHz) | $f_M^s$ (GHz) | $m_{eff}^s$ (pg) | $g_{om}^e$ (GHz/nm) | $g_{om}^s$ (GHz/nm) | $g_0^e$ (MHz) | $g_0^s$ (MHz) |
|---|---|---|---|---|---|---|---|
| 0.98 ± 0.05 | 1.39 | 1.39 ± 0.05 | 2.2 | 485 | 693 | 0.8 | 1.14 |
| 1.04 ± 0.05 | 1.36 | 1.37 ± 0.05 | 1.4 | 277 | 396 | 0.58 | 0.84 |
| 1.03 ± 0.05 | 1.34 | 1.34 ± 0.05 | 2.4 | 429 | 613 | 0.7 | 1.0 |

FIG. 1. (Color online) (a) Scanning electron microscope (SEM) view of a GaAs disk (1 μm diameter and 200 nm thickness) suspended on an AlGaAs pedestal. (b) Electric field intensity $|\mathbf{E}(\mathbf{r})|^2$ of the optical mode (p=1, m=7). (c) Deformation profile of the M = 0 mechanical breathing mode. (d) Schematics of the near-field optomechanical spectroscopy experiment. The linear polarization is selected by a fiber polarization controller (FPC) and a half-wave plate. PD stands for photodetector.

FIG. 2. (Color online) (a)-(c) Normalized optical transmission spectrum of optical resonance (p=1, m=7) measured at low and high optical power in three disks. (d) The WGM wavelength as a function of the disk radius. (e) Relative optical resonance shift Δλ as a function of the total transmitted optical power. (f) The temperature change as a function of normalized resonance shift $\Delta\lambda/\lambda_o$ due to thermal-optic effect.

FIG. 3. (Color online) (a)-(c) The RF-spectrum of the M = 0 breathing mechanical mode from each disk taken in ambient conditions. (d) A calibrated Brownian motion spectrum of disk 1.

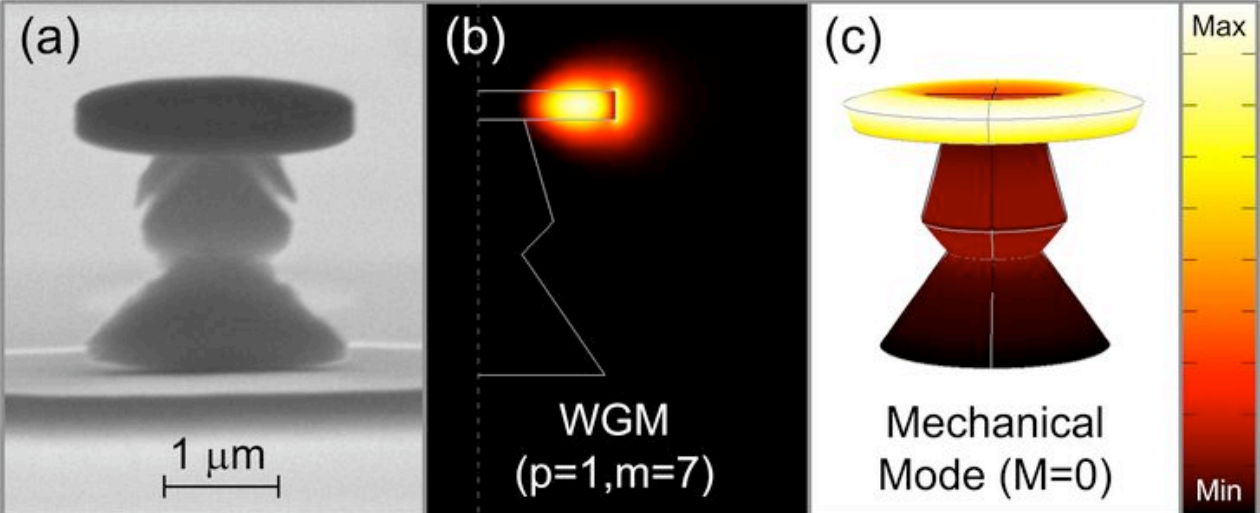
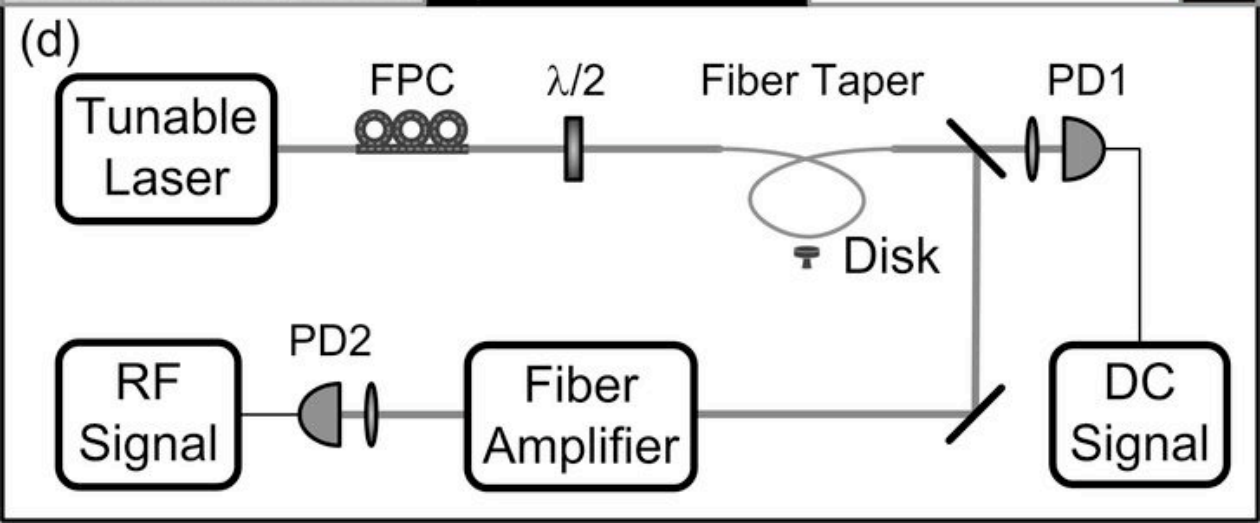

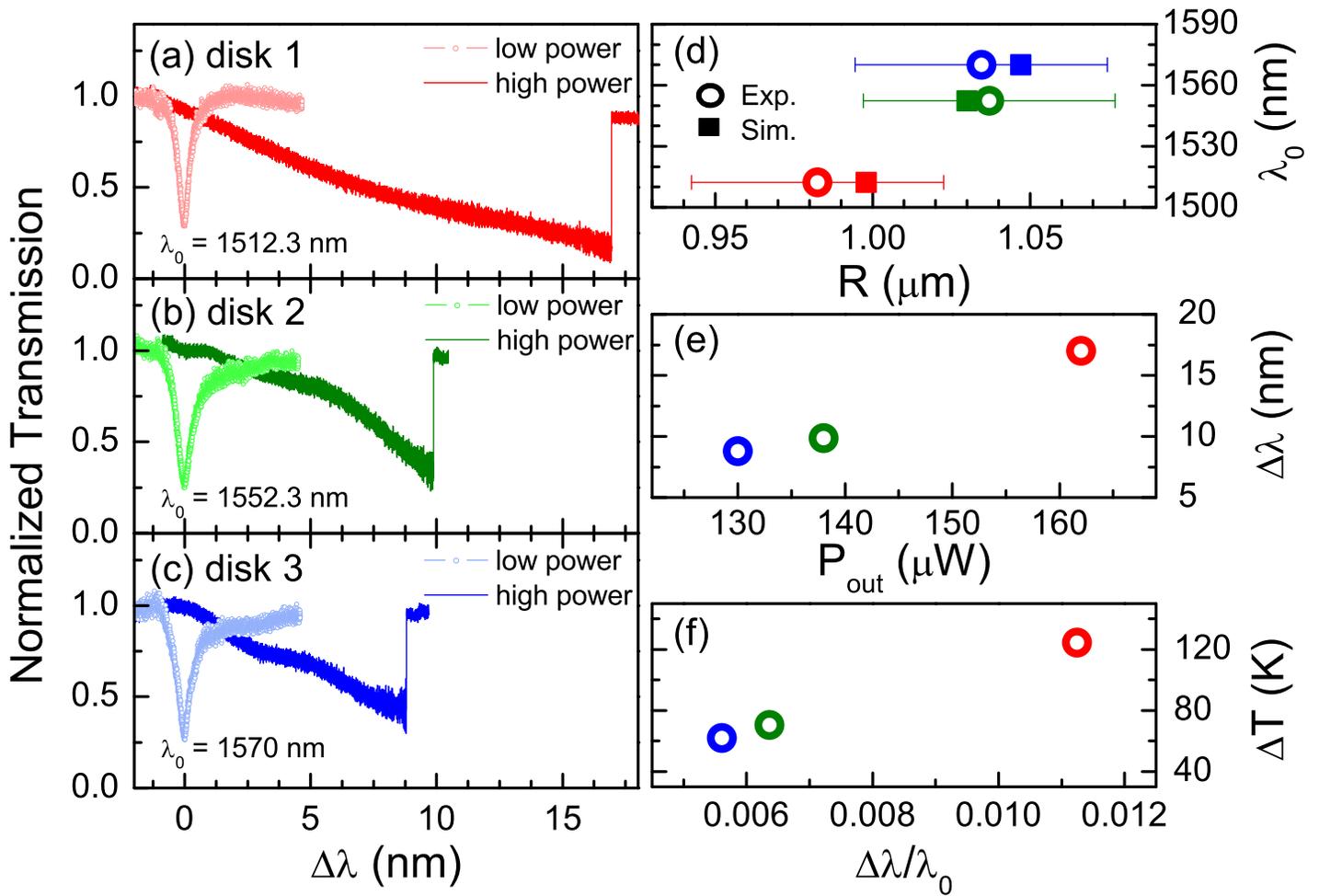

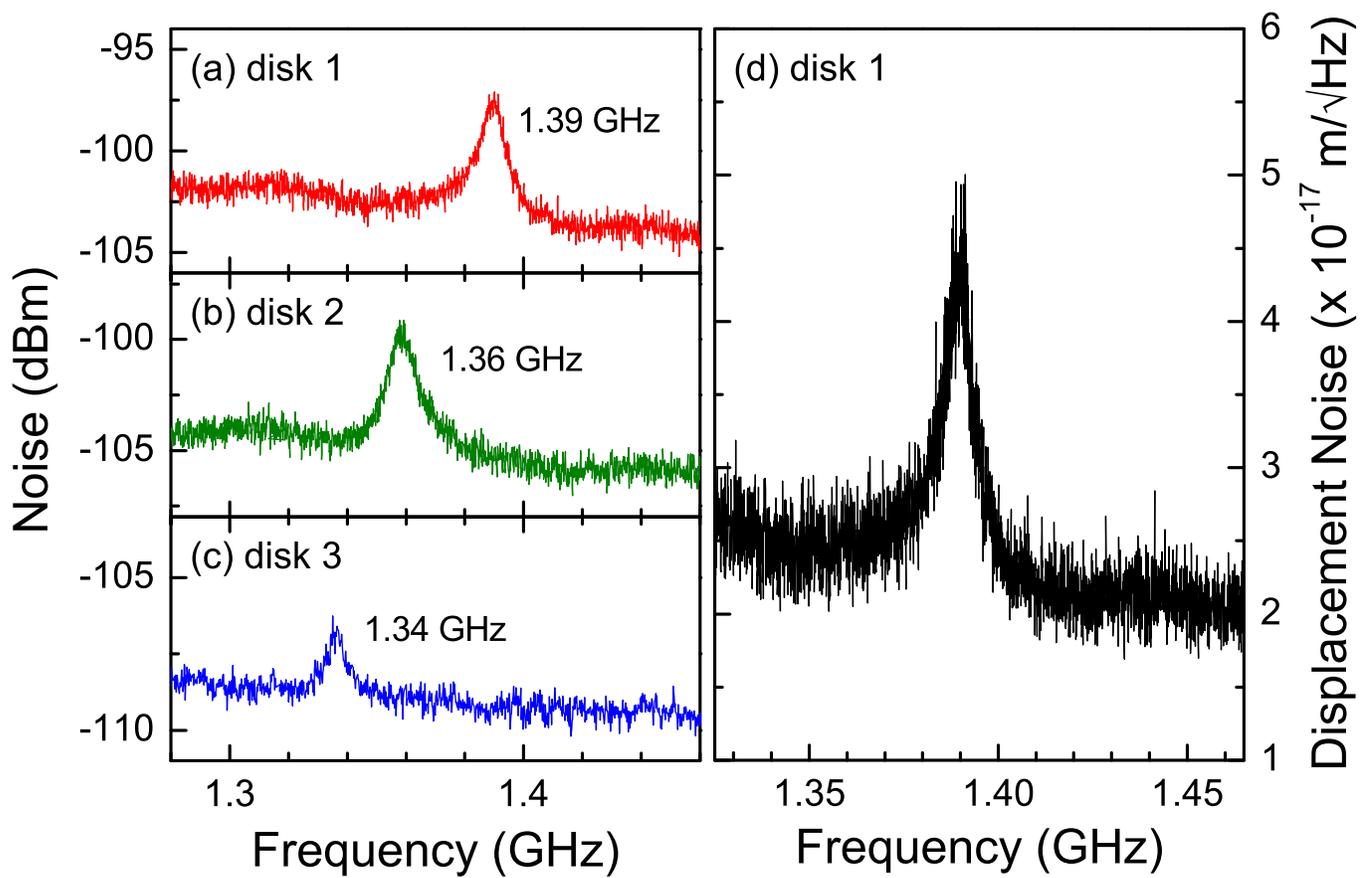